\documentclass[aps,twocolumn,superscriptaddress,altaffilletter,lengthcheck, tightenlines,showpacs,showkeys]{revtex4}

\newcommand{\ben}{\begin{eqnarray}}
\newcommand{\een}{\end{eqnarray}}
\newcommand{\be}{\begin{equation}}
\newcommand{\ee}{\end{equation}}
\newcommand{\ba}{\begin{eqnarray}}
\newcommand{\ea}{\end{eqnarray}}

\newcommand{\ga}{\gamma}
\newcommand{\ro}{\rho}

\newcommand{\bn}{\begin{equation}\label}

\usepackage[dvipdf]{epsfig}
\usepackage{color}

\begin{document}

\title{Cosmological perturbations in transient phantom inflation scenarios}

\author{Mart\'{\i}n G. Richarte}\email{martin@df.uba.ar}
\affiliation{Departamento de F\'isica, Universidade Federal do Paran\'a, Caixa Postal 19044, 81531-990 Curitiba, Brazil}
\affiliation{Departamento de F\'isica, Facultad de Ciencias Exactas y Naturales,
Universidad de Buenos Aires, Ciudad Universitaria 1428, Pabell\'on I,  Buenos Aires, Argentina}
\author{Gilberto M. Kremer}\email{kremer@fisica.ufpr.br}
\affiliation{Departamento de F\'isica, Universidade Federal do Paran\'a, Caixa Postal 19044, 81531-990 Curitiba, Brazil}
\bibliographystyle{plain}

\begin{abstract}
We present a model of inflation where  the inflaton is accommodated as a phantom field which exhibits an initial transient pole behavior and then  decays into a quintessence field which is  responsible for a radiation era.  We must stress that the present unified model only deals with a single  field and that the transition between two eras is achieved in a smooth way, so the model does not suffer from the eternal inflation issue.  We  explore the conditions for the crossing  of the phantom divide line within the inflationary era  along with the  structural stability of several critical points. We study the behavior of the phantom field within the slow climb approximation along with  the  necessary conditions to  have sufficient  inflation. We also examine the model at the level of  classical perturbations  within the Newtonian gauge and determine the behavior of the gravitational potential, contrast density  and perturbed field near the inflation stage and the subsequent radiation era. 
 \end{abstract}
\vskip 1cm
\keywords{pole inflation, phantom field,  graceful exit, quintessence field, radiation,  perturbation theory.}

\date{\today}
\maketitle
\section{Introduction}  
The standard inflationary paradigm is a theoretically appealing proposal invented to solve
many problems which arise in the hot big bang model \cite{guth}. Although the latter one  was extremely successful
in explaining the observational evidence of an expanding universe, it had several problems which motivated the first
inflationary models. For instance, the inflationary model was able to explain the flatness of the universe at large scale
without using  a fine-tuning mechanism  to set the initial condition in the very early universe.  The hot big bang suffers
from the horizon problem, which means that different regions of the sky could not have been causally connected at the time of the hot
big bang, however,  the almost uniform CMB temperature across the sky is a bullet proof of the large scale homogeneity in the universe \cite{liddle1a}.
Another issue that arises in the hot big bang model refers to the amount of monopoles observed in the universe. 
According to  the grand unified theories of particle
physics the hot big bang would produce a vast number of magnetic monopoles.
The predicted length scales of their creation is such that the universe would be
large enough at the GUT energy scale to contain a significant number, yet a
magnetic monopole has never been observed so far in nature \cite{liddle1b}. The inflationary scenario dictates that during some short epoch in the very early universe the non-zero vacuum energy density of some unknown field dominated the energy density
of all other forms of energy, such as matter or radiation. In the simplest case,  inflation
is caused by a cosmological constant. A more complete scenario is inflation driven by a
slowly rolling scalar field in a potential well. During the inflationary phase the scale factor of
the universe grew exponentially so that initially small patches of space could have been
stretched greater than the current observable universe \cite{liddle2}. One vital piece of this
mechanism is the capability of explaining the origin of the primordial seeds for the cosmic large scale structure and
fluctuations in the cosmic microwave background \cite{liddle3}, \cite{muka1}, \cite{muka2}.

As was mentioned earlier, one of the most simplest scenario is a scalar inflaton field (associated with spin 0 particle) endowed with the property of having
a sufficiently negative pressure to violate the strong energy condition. Moreover, the inflationary phase
begins near the maximum of a flat potential; then  the inflaton slowly rolls down the potential and finally the end of inflation is  marked by 
oscillations at the minimum. This cornerstone of modern cosmology has been the subject of much investigation during the last decades \cite{liddle3}. 

There is a vast landscape of cosmological models where inflation can be achieved, for instance, vector model, extended gravity theory, gauge fields,  coupled scalar fields, and many others   \cite{liddle2}, \cite{liddle3}, \cite{muka1}, \cite{muka2}. However, we are going to focus our research on a recent proposal dubbed ``phantom inflation'' \cite{ph1}, \cite{ph1a}-\cite{ph1e},  \cite{ph2}. A somewhat natural manner to realize this scenario is by considering a self-interacting scalar field with a negative kinetic energy in the Lagrangian \cite{ph3}, \cite{ph4}. The reason for examining such kind of inflation is threefold. First, phantom-like fields can be found in several types of well-accepted theories, for instance, k-essence models \cite{ke1}, \cite{ke2}. Besides, the current accelerating state of the universe can be properly described in terms of a phantom field with a negative equation state and this possibility cannot be discarded by the recent data extracted from Planck mission \cite{planck}. Third, the phantom inflation model has quite distinctive feature in relation with the standard model, which makes it an interesting alternative model. To be more precise, one peculiar  fact is that the old minimal conditions to achieve inflation (the so- called slow-roll approximation) must be replaced by a slow-climb mechanism provided the self-interacting potential must exhibit a maximum in order to avoid a fatal primordial big rip.  Regarding the  curvature power spectrum, the phantom inflation can also lead to a nearly scale  invariant one, but its tensor perturbation is characterized by a blue tilt instead of a red one \cite{ph1}. Nevertheless, one of the most critical shortcoming of the aforesaid model is the possibility of extending in time forever, thus not allowing for a graceful exit from the inflation era. This point  was carefully analyzed first by  Piao and  Zhang \cite{ph1}. These authors employed  a hybrid extension of the phantom model by including an extra-normal scalar field,  indeed, the coupling between them  was designed  for the phantom field to attain a slow-climb phase and then decay (for certain critical mass scale) into radiation due to the presence of the normal scalar field \cite{ph1}. The existence of a graceful exit from a phantom-like inflationary era was also explored in more detail in the context of non-minimally coupled gravity theory \cite{ema};  Feng et al. included an extra ${\cal F(R)}\phi^2$ coupling in the phantom inflation model to achieve such goal also, $\phi$ being the phantom field and ${\cal F(R)}$ a generic function of Ricci scalar \cite{ema}. Interestingly enough,  the new slow-climb conditions were derived and  the general  necessary conditions to avoid eternal phantom inflation were presented as well, finding that finite phantom inflation can be achieved without invoking new extra fields \cite{ema}, \cite{tech}. 

One of our aims in the present work is to show the possibility of having a phantom inflation era where the universe exhibits a pole-like behavior in the cosmic time and then transits to another matter phase without evoking new extra fields. In doing so,  the slow-climb conditions are satisfied during the short-inflation era. In order to achieve this goal, we are going to parametrize the scale factor  with a double power law in terms of the cosmic time. With this proposal at hand, we will be able to reconstruct both the kinetic and the potential energy in terms of the cosmic time; in particular, we are going to obtain  the shape of the potential in terms of the phantom scalar field in the limiting cases of large and small times. We  examine the classical perturbation of this model, calculating the leading behavior of the gravitational Newtonian potential, the curvature perturbation along with the behavior of the perturbed phantom field.   Such findings show that the extended model is free from the eternal inflation issue but more importantly the transition between the two eras  is achieved in smooth way  and without the inclusion of new-extra scalar field.  

The paper is organized as follows. In Sec. II, we present the  phantom inflation model and discuss its main ingredients. In Sec. III, the conditions for the crossing  of the phantom divide line within the inflationary era  are discussed along with the  structural stability of several critical points. Sec. IV is devoted  to the analysis of the slow-climb approximation. In Sec. V the phenomenology of super-inflation is discussed.  In Sec. VI the  analysis of the classical perturbation  is performed. The conclusions are stated  in Sec. VII.

\section{Transient phantom inflation}
Our starting point is to consider a phantom inflaton field  minimally coupled to gravity.  We then use as our action  
\bn{a}
{\cal S}=\int{d^4{x}\sqrt{-g} \left[\frac{{\cal R}}{2} +\frac{1}{2}g^{\mu\nu}\partial_{\mu}\phi\partial_{\nu}\phi -V(\phi) \right]},
\ee
where $\phi$ stands for the phantom (inflaton) field, $V$ represents the self-interacting potential and ${\cal R}$ is the Ricci scalar constructed from the metric theory. Throughout the article we work with natural (geometrical) units such that $M^{2}_{\rm p}=8\pi G=1$ and we adopt the metric signature $(-,+,+,+)$.
Taking into account (\ref{a}) we can  derive the so called the Einstein field equations where  the energy-momentum  is of  the form 
\bn{tmn}
T^{\phi}_{\mu\nu}=-\frac{2}{\sqrt{-g}}\frac{\delta {\cal L}^{\phi}}{\delta g^{\mu\nu}}=-{\cal L}^{\phi}_{,X}\partial_{\mu}\phi\partial_{\nu}\phi + g_{\mu\nu} {\cal L}^{\phi}.
\ee
Here ${\cal L}^{\phi}=X-V$ stands for the phantom Lagrangian (\ref{a}), $X\equiv (1/2)g^{\mu\nu}\partial_{\mu}\phi\partial_{\nu}\phi<0$, and the suffix $_{,X}$ represents a partial derivative with respect to $X$. Varying with respect to the phantom field it leads to a modified Klein-Gordon equation
\bn{kg}
\frac{1}{\sqrt{-g}}\partial_{\mu}(\sqrt{-g}\partial^{\mu}\phi)-V,\phi=0.
\ee

We are going  to work in the spatially flat Friedmann-Robertson-Walker (FRW) metric with cosmic scale factor $a(t)$, namely, the metric is  $g_{\mu\nu}={\rm diag}[-1,  a^{2}(t)\delta^{i}_{j}]$ with $t$ the cosmic time and $\delta^{i}_{j}$ the delta Kronecker tensor. The energy-momentum tensor of the phantom field (\ref{tmn}) can be recast  in terms of the one associated with a perfect fluid. Assuming an homogeneous phantom field,  its pressure and energy density  are given by
\bn{ro}
\ro_{\phi}=-\frac{\dot{\phi}^2}{2}+ V(\phi),
\ee
\bn{pe}
P_{\phi}=-\frac{\dot{\phi}^2}{2}- V(\phi),
\ee 
where the dot stands for derivatives with respect to the conformal time.  The  Friedmann  and the Raychaudhuri equations are given by  
\bn{fri}
3{ H }^2= \ro_{\phi},
\ee
\bn{ray}
\frac{\ddot{a}}{a}={H }^2+ {\dot{ H }}=-\frac{1}{6}(\rho_{\phi}+3P_{\phi}),
\ee 
where the  Hubble parameter is  defined as $H =\dot{a}/a$. Eq. (\ref{ray}) tells us that an accelerated phase can be obtained if $\rho_{\phi}+3P_{\phi}<0$, which is equivalent to $-2({\dot{\phi}}^2+V)<0$. As one could expect, the weak energy condition is ruled out, that is, $\rho_{\phi}+P_{\phi}=-{\dot{\phi}}^2<0$. It should be remarked that the local energy conservation
\bn{ce}
 \dot{\rho}_{\phi}+3{H}(\rho_{\phi}+P_{\phi})=0
\ee
is not altered within the phantom inflation provided it follows from Bianchi identity ( i.e. $T^{\mu\nu}_{~~~;\mu}=0$) regardless of the specific form of energy-momentum that appears in the r.h.s of Einstein's equations. Moreover, Eq. (\ref{ce}) is equivalent to the modified KG:
\bn{kg}
\ddot{\phi} +3 H\dot{\phi} -V_{,\phi}=0.
\ee
 
At this point we are going to reconstruct  the potential energy and the kinetic energy from a given scale factor so  it is essential to show that procedure is easy to follow but also it will give us all the physical information  encoded in the phantom inflationary scenario at the background level.  Combining (\ref{fri}) with (\ref{ce})   we can obtain some formal expressions for the potential and kinetic term in terms of kinematic quantities:  

\bn{kt}
{\dot{\phi}}^{2}=2\dot{ H},
\ee
and
\bn{pot}
V=\dot {H}+3{ H}^2. 
\ee
Knowledge of the evolution of  $V$ and ${\dot{\phi}}^2$  requires to solve the Friedmann equation (\ref{fri}) along with the conservation equation (\ref{ce}) consistently. In general, such task is not easy to tackle provided the complexity and non-linearity of the above system of equations. In fact, the general exact solution for the scalar field is only known for an exponential potential \cite{luis1}, \cite{luis2}. Our aim here is not to provide the general exact solution but to restrict our attention to a new (but quite general in some sense) kind of scale factor, which is an extended version of the power law. Given the scale factor $a(t)$, we will proceed to re-build the potential at the background level. Without further delays, we parametrize scale factor as the product of two power laws:
\bn{at}
a(t)=a_{i}t^{\gamma}\left[1+\frac{t}{w} \right]^{\beta-\gamma}, 
\ee
where $a_{i}$ is a reference scale factor while $\beta, \gamma,$ and $w$ are parameters of the double power-law model.  Inflation will last for a period of time  $t  \in [ 0^{+}, t_{c}]$  and after $t> t_{c}$ the universe enters into non-accelerated phase. The physical motivation for   this ansatz is connected with the string theory, where  an exact solution with a pole behavior was reported \cite{string}. Our purpose is to generalize this result within the framework of GR, giving simultaneously  a new mechanism within the phantom scenario in which  pole inflation decays toward a matter era at late times without introducing extra fields into the model. To make things clearer, we need to analyze the limiting behavior of the scale factor (\ref{at}) for large and small values of the cosmic time.  Let us first  consider the case of small cosmic time. Eq. (\ref{at}) becomes 
\bn{a0}
a \simeq a_{i} t^{\gamma}[1+ \frac{t}{w}(\beta -\ga)],
\ee
so the Hubble function is $ H\simeq \ga t^{-1}$ whereas the acceleration term approaches  $\ddot{a}/a \simeq  \ga (\ga -1)t^{-2}$. Bearing in mind the possibility of having an early accelerated era, we have two choices for $\ga$; it could be $\ga>1$ or $\ga<0$ \cite{f}. Taking into account the behavior of $H$ and $\ddot{a}/a$, we can conclude that $\ga>1$ is featured by accelerated contraction ($\dot{H}<0$ and $\ddot{a}/a>0$) but the other case with $\ga<0$ describes a phase of accelerated expansion driven  by a phantom field with negative pressure. In fact, the aforesaid situation is more interesting provided  the universe is super-accelerated  because $ \dot{H}>0$ and $\ddot{a}/a>0$, so the scalar curvature  of the space-time is growing in the limit $t \simeq 0^{+}$ \cite{f2}. Calculating the leading term in the energy density (\ref{fri}), we arrive at 
\bn{dt}
\rho_{\phi}(t)=3\rho_{i} \ga^2  t^{-2}. 
\ee 
Now, we give a physical interpretation of the dimensionless constant $w$. As is well known, during inflation, the scale energy is fixed by the inflaton. This equivalent to say that the energy during inflation should scale as $\rho_{\inf} \simeq \Lambda^{4}$, $\Lambda$ being a typical scale related with the Planck's mass. Identifying the constant in (\ref{dt}) with the cutoff $\Lambda$, we obtain that $\rho_{i}=\Lambda^4/(3\ga^2)$,  being $\rho_{i}$  an initial energy scale; such value could  be related with the critical density of the universe today.  A better physical insight can be gained by calculating the leading term  in  the pressure given by $P_{\phi}=-[2\dot{H} + 3H^2]$. It turns out that the pressure is  given by
\bn{pt}
P_{\phi}(t)=\rho_{i} \ga (-3\ga +2) t^{-2}. 
\ee 
Plugging (\ref{dt}) into (\ref{pt}), we see that the phantom field behaves as a dressed cosmological constant provided the new equation of state is $P_{\phi}=-\mu \rho_{\phi}$ with $\mu=(1+ 2/3|\ga|)>0$ for $\ga<0$.  Eq. (\ref{kt}) tells us that the leading term in the kinetic energy is ${\dot {\phi}}^2(t) \simeq -2\ga t^{-2}$; then the phantom field can be recast as $\phi=\phi_{0}-\epsilon_{\pm} \sqrt{2|\ga|}\ln t$ with $\epsilon_{\pm}=\pm 1$. During this regime $\rho_{\phi} \simeq V$, so the potential can be written as 
\bn{pe1}
 V=V_{1}\exp[\lambda_{1} \epsilon_{\pm} (\phi -\phi_{0})],
\ee
where $V_{1}=V_{i}\ga(3\ga-1)$ and $\lambda_{1}=(2/|\ga|)^{1/2}$ \cite{f}. An interesting fact is that the potential (\ref{pe1}) can be the exponential one but also its counterpart; a negative exponential. Notice that the positivity of $V_{1}$   and the reality of $\lambda_{1}$ are both  guaranteed provided  the condition $\ga<0$ holds for pole inflation (or super-inflation).

To this end, we need to explore the physical outcomes for considering large cosmic time. The scale is given by 
\bn{a2}
a(t) \simeq a_{i}\frac{t^{\beta}}{w^{(\beta -\gamma) }}.
\ee
From (\ref{a2}), we obtain $H=\beta t^{-1}$ along with $\ddot{a}/a \simeq \beta(\beta-1)t^{-2}$. A critical issue in phantom inflationary model is to ensure a graceful exit from that primordial era, that is,  allowing a transition toward a non-accelerated era such as radiation or matter. Therefore, a non-accelerated regime after phantom inflation is achieved if $0<\beta <1$ provided $\ddot{a}<0$  and $\dot{H}= -\beta t^{-2}<0$. On the other hand, the density  and pressure associated with the phantom fields are given by 
\bn{dl}
\rho_{\phi} \simeq 3\rho_{i}\beta^2 t^{-2},
\ee

\bn{pl}
P_{\phi} \simeq p_{i} \beta(2-3\beta)t^{-2}.
\ee
Identifying the energy density (\ref{dl}) in terms of the scale factor $\rho_{\phi} \simeq 3\rho_{i}\beta^2 w^{-2(1+|\ga|/\beta)} (a/a_{i})^{-2/\beta}$ with a radiation-like matter $\rho =\rho_{r}(a/a_{i})^{-3\ga_{r}}$ ($\ga_{r}$ being the barotropic index for radiation matter), one obtains $\rho_{r}=3\beta^{2}\rho_{i}[w^{-2(1-\ga/\beta)}]$, so $w=[3\beta^2 (\rho_{i}/\rho_{r})]^{m}$ with $m=\beta/2(\beta +|\ga|)>0$. To have exact radiation $\beta$ should be fixed at the $1/2$ value. Furthermore, combining (\ref{dl}) and (\ref{pl}), we find that the equation of state is $P_{\phi} \simeq (\frac{1}{3}+ \Delta)\rho_{\phi}$; so it is quite similar to radiation as long as $0<\beta<(2/3)$.  One can estimate the deviation from a radiation matter encoded in the value of $|\Delta|=(2/3)|(1/\beta)-2|$. The potential in terms of the cosmic  time can be recast  as $V \simeq V_{i} \beta (3\beta-1)t^{-2}$. However, the identification by means of the phantom field is  not sot evident because the phantom field becomes oscillatory, namely, $\phi=\phi_{0}+\epsilon_{\pm} i\sqrt{2\beta}\ln t$ and the potential is represented as $V \simeq V_{i} \beta (3\beta-1) e^{i\epsilon_{\pm}\lambda_{2}(\phi -\phi_{0})}$ with $\lambda_{2}=\sqrt{2/\beta}$ as well. 

\section{Equation of state and the structural stability }
Let us begin this section by mentioning some important comments regarding the  behavior of equation of state associated with the inflationary phantom model. 
As is well known,  phantom inflation should have a graceful exit   along with  a  crossing  of the phantom divide and only then should exit the accelerated stage \cite{vikman}. Some authors   found that a  k-essence  associated with a scalar field with the quadratic kinetic term   cannot achieve the crossing of the phantom divide in a physically plausible way \cite{vikman}. So,  we are going to see what happens with our model and the crossing of the phantom divide.  
To do so,  we need to calculate the  equation of state parameter defined as ${\cal W}(t)=-1-2\dot{H}/3H^{2}$. In doing so,  we use the scale factor (\ref{at}) along with its derivatives to  express  the  equation of state parameter in the following form:
\bn{esp}
{\cal W}(t)=-1+ \Big(\frac{\beta t^2+2\gamma wt+ \gamma w^2}{\beta^2t^2+2\gamma\beta wt+ \gamma^2w^2}\Big).
\ee
Equation (\ref{esp}) tells us that  near the origin the value of the equation of state parameter does not depend on other parameters than $\gamma$, namely ${\cal W}(t\rightarrow 0^{+})=-1+2/3\gamma$.  In order to have an initial inflationary phase one has to impose  $\gamma<0$  and therefore  ${\cal W}$ is clearly negative.  In the opposite case, we obtain   ${\cal W}(t\rightarrow \infty)=-1+2/3\beta>0$ for $0<\beta< 2/3$; once again the non-accelerated phase is controlled by the parameter $\beta$ only. For instance, one can have  a matter dominated era  by choosing $\beta=1/3$ or a radiation phase for $\beta=1/2$ and both cases satisfy the condition  ${\cal W}(t\rightarrow \infty)>0$. From the latter results, it is clear that there is a smooth transition between an accelerated phase toward a non-accelerated final stage, which is one of the main purposes of the present work.  

Another fact that can be noticed from Eq. (\ref{esp}) is that the crossing of the phantom divide, namely ${\cal W}(t_c)=-1$,  it can  be achieved when $\dot{H}$ vanishes. The latter fact  is equivalent to the requirement that the quadratic equation in Eq. (\ref{esp})  must have at least one zero. In order to gain some insights about that point, we are going to choose $w=1$ and $\gamma=-100$ for the rest of the discussion. Nevertheless,  this choice is quite generic and we are not losing generality in the process of analyzing the crossing divide.  We are going to consider  two physical cases, the one with $\beta_{m}=1/3$ associated with a final era dominated by matter and another case with $\beta_{r}=1/2$ which corresponds to a radiation dominated era.  For  $\beta_{m}=1/3$, we see that the condition $\dot{H}=0$ is obtained twice, indicating that the universe  crosses  the phantom divide twice also. To be more precise,  the numerical analysis shows that  the condition ${\cal W}=-1$ is reached first at $t^{m}_{-} \simeq 55.01{\rm seg}$   and later at $t^{m}_{+} \simeq 544.94{\rm seg}$. Of course,  one expects that the universe leaves the accelerated initial stage after having crossed the divide line so  the time at which the acceleration is zero, called $t_{c}$, must be greater than  $t^{m}_{+}$. In the  $\beta_{m}=1/3$ case we find that $t_{c}=668.04 {\rm seg}>t^{m}_{+}$. The same situation happens for the case with  $\beta_{r}=1/2$ in the sense that the universe crosses the phantom divide line twice and the last crossing occurs before the universe enters the non-accelerated phase. However, the universe dominated by radiation reaches the last crossing of the phantom divide faster than the universe dominated by matter. In order to really understand how our model does differ from previous proposals based on phantom models \cite{vikman}, we need to explain how the final stage can be achieved in terms this ``phantom'' field and  we will return to this point later.   

We are going to investigate now the existence of asymptotic power-law solutions by means of a structural stability analysis, in other words, we are going to unveil the asymptotic nature of potentials allowing such solutions. To this end we introduce the equation of state parameter ${\cal W}=p/\rho$, and from the definitions above (\ref{ro} -\ref{pe}) and the Einstein equations (\ref{fri}) along with the conservation equation (\ref{ce}) one gets the dynamical equations for ${\cal W}(t)$, namely
\bn{masterw}
\dot{{\cal W}}= ({\cal W}-1)\Big[\frac{\dot{V}}{V}+ 3H ({\cal W}+1)\Big]. 
\ee
In addition, one has to use  the expression $-2V=({\cal W}-1)\rho$ along with Eq. (\ref{kg}) to obtain the above master equation (\ref{masterw}). The reason for studying the stability of power-law solution in terms of the equation of state parameter is that the universe has a scale factor given by a power-low solution in the initial and final stages. Most importantly, the physical motivation for choosing a master equation with the ${\cal W}$  as a physical variable is  that  the universe has a constant equation of state parameter at the beginning of its evolution. Clearly, the solution ${\cal W}_{m}=1$ represents an equilibrium
point, but it corresponds to non-accelerated expansion, specifically, $a \propto t^{2/3}$ .  Let us require then that Eq. (\ref{masterw}) admits another equilibrium point with ${\cal W}={\cal W}_{c}$,  representing in consequence a solution with $a \propto t^{2/3({\cal W}_{c}+1)}$. In particular, a solution representing a phantom early-time repeller would be characterized by a constant and negative value of ${\cal W}_{c}$. It is important to remark that contrary to what happens with a phantom field used to describe an era dominated by dark energy at late times \cite{ito}, the initial inflationary stage associated with the critical point ${\cal W}_{c}$ must be a repeller; otherwise the universe cannot leave the aforesaid phase and the transition to another era would not be  possible at all.  Below, we will show that the requirement of the existence of such an equilibrium point restricts the functional form of the $V (\phi)$ potential. Then we formulate a structural stability analysis of the equation governing the evolution of the parameter ${\cal W}$. This is equivalent to imposing the
asymptotic condition on the potential $\dot{V} + 3 ({\cal W}+1) HV = 0$, which can be integrated to give $V = V_{0} a^{-3({\cal W}_{c}+1)}$ with $V_{0}$ a positive integration constant. This finding about the form of the potential is completely consistent with our previous result (\ref{dt}) provided $\rho \simeq V(t)$. By choosing this potential the asymptotic regime of ${\cal W}$ is governed by the equation
\bn{masterw2}
\dot{{\cal W}}=3H({\cal W}-1)\big[{\cal W}-{\cal W}_{c}\big]. 
\ee
Let us now define ${\cal W} = {\cal W}_{c} + \epsilon$ with $0<\epsilon\ll 1$ and recast Eq. (\ref{masterw2}) in
terms of this new definition. By expanding the r.h.s of this equation up to order $\epsilon$, one obtains $\dot{\epsilon}=3H\epsilon ({\cal W}_{c}-1)$. Then,  
for initially contracting universes $(H < 0)$, the solution ${\cal W} = {\cal W}_{c}$ is a repeller  provided ${\cal W}_{c}<0$, which is on the other hand
our working hypothesis for having an inflationary stage ($\ga<0$). As  has been shown,  the initial stage must be associated with a repeller in order to have a transition to a second non-accelerated phase otherwise a graceful exit cannot be guaranteed.  We are in position to investigate the stability of the solution  ${\cal W}_{c}(t_{bw})=-1$ associated with the last transition of phantom divide line. If we impose the fulfillment of Eq. (\ref{masterw2})  around $t_{bw}$ one obtains $\dot{V} /V =0$ with $V = V_{0}$ a constant. Now defining ${\cal W} = -1+ \epsilon$ with $0<\epsilon<<1$ and using again (\ref{masterw2}) it follows that  the master equation becomes $\dot{\epsilon}=3H(\epsilon-2)^2\simeq 12 H(1-\epsilon)>0$.  So for the inflationary phase with $H>0$ leads to  an unstable transition at ${\cal W}_{c}=-1$, however, such transition is harmless because the universe must cross the divide line before enters the non accelerated era. If the transition were charaacterized by an attractor point (stable) the  phantom inflation would not be able to cross the phantom divide and therefore it would remain in that attractor forever, indicating  that the model is not physically interesting provided cannot guaranteed the graceful exit from inflation along with the crossing of the phantom divide before moving forward the second stage \cite{vikman}.  At this point, it should be remarked that the analysis of instability  associated with the crossing of the divide line performed in Ref. \cite{ito} does not hold in our case. The main reason is that in our inflationary  stage the field is real and does not  go like $\phi=t$, so their working hypotheses are different from the ones used in our work. Furthermore, we have shown that the structural stability of the power-law solution imposes  the form of the potential and  consistently the Friedman equation along with the Klein-Gordon equation determines the asymptotically shape of the phantom field. Nevertheless, one can consider both analyses as complementary ones given the fact that they describe different cosmological scenarios.

Let us end this section by showing that there is a symmetry in the proposed model or Lagrangian which can help us to understand the final stage where the model is accommodated as a scalar field which describes a radiation- or mater-like era. The Lagrangian is given by ${\cal L}=(1/2)\partial_{\mu}\phi\partial^{\mu}\phi-V(\phi)$. We begin by  considering the existence of a symmetry which modifies the field in a global phase, namely  $\phi \rightarrow i \varphi$ and $\partial^{\mu}\phi \rightarrow i \partial^{\mu}\varphi$,  so the transformed Lagrangian becomes  ${\cal L}'=(1/2) i^{2}\partial_{\mu}\varphi\partial^{\mu}\varphi-V(i\varphi)$. Therefore, a  realization of the symmetry only requires potentials which remain real under the transformation of the field $\phi \rightarrow i \varphi$ with $\varphi$ real. The existence of this kind of symmetry is equivalent to say that the original phantom field becomes a quintessence field with a normal kinetic energy. It is exactly because of the aforesaid property that the model can describe the initial era with a phantom Lagrangian and a final era with a quintessence Lagrangian. To be more precise, during the initial era it behaves as a phantom (real) field and then it becomes an pure imaginary field which leads to another scalar field with the normal kinetic energy and therefore it is suitable to account for  a radiation  era or a matter-like stage. One way to understand this point is with the help of Eq. (\ref{kt}), during the first accelerated era $\dot{H}>0$ so $\dot{\phi}^2$ remains positive and the model is associated with a phantom density, namely $\rho=-(\dot{\phi}^2)/2+V(\phi)$. When the universe enters the non-accelerated phase, this means that $\dot{H}<0$ and by virtue of Eq. (\ref{kt}) it leads to $\dot{\phi}=i \sqrt{|2\dot{H}|}=i\dot{\varphi}$ then the energy density becomes $\rho'=+(\dot{\varphi}^2)/2+V(\varphi)$. The previous statement is consistent with our findings reported at the end of Sec. II, where we obtained  $\varphi=\varphi_{0}+\epsilon_{\pm} \sqrt{2\beta}\ln t$ and the potential becomes real in terms of the $\varphi$ field, namely $V \simeq V_{i} \beta (3\beta-1) e^{\epsilon_{\pm}\lambda_{2}(\varphi -\varphi_{0})}$.

\section{Slow-climb approximation}
The condition for  having inflation is $\rho_{\phi} + 3P_{\phi} =-2({\dot{\phi}}^2+V)<0$.  The previous condition should be valid sufficiently long time  in order to make the universe sufficiently flat. The standard approximation technique for analyzing inflation is the slow-roll approximation \cite{liddle2}, however, this approach must be re-considered in the phantom scenario providing the negativity of kinetic energy term. 
In turns out that a viable phantom inflationary model must fulfill  the following conditions,  called  slow-climb criteria \cite{ph1}, \cite{ph2}:

\begin{itemize}
  \item $|\dot{H}|\ll H^{2}$. This is equivalent to ensure $|\ga|\gg 1$.
  \item $|\ddot{\phi}|\ll3|H\dot{\phi}|$ or $3|\ga|\gg1$.
  \item $2V\gg|-\dot{\phi}^2|$ or $3|\gamma|\gg1$.
\end{itemize}
Taking into account the above results, we can roughly estimate the value of the slow-climb (SC) parameters in the primordial inflationary era.  The SC parameters can be written as \cite{liddle2}, \cite{ph1}, \cite{ph2}
\bn{e1}
\epsilon_{\rm ph} = -\frac{\dot{H}}{H^2},
\ee
\bn{e2}
\delta_{\rm ph}=-\frac{\ddot{\phi}}{H\dot{\phi}},
\ee
and the difference between (\ref{e1}) and (\ref{e2}) is
\bn{e3}
\eta_{\rm ph}=(\epsilon_{\rm ph}-\delta_{\rm ph}).
\ee
Eqs. (\ref{e1})-(\ref{e3}) tell us that $\epsilon_{\rm ph}=1/\ga$, $\delta_{\rm ph}=-1/\ga$ and therefore $\eta_{\rm ph}=2/\ga \neq 0$. Clearly, the condition $|\ga|\gg1$ implies the smallness of the slow-climb parameters. Additionally, similar results can be obtained  when  the slow-climb parameters are parametrized by the smallness of the potential slope and its concavity \cite{liddle2}. Then  we arrive at  $\epsilon_{\rm Vph}= (1/2)(V_{,\phi}/V)^2=1/|\ga|\ll1$ along with $\eta_{\rm Vph}=V_{,\phi\phi}/V=2/|\ga|\ll1$.  

In a similar fashion, we can inspect the viability of the SC approximation by including higher terms of the hierarchy (cf. \cite{liddle2}). For instance, the square and cubic SC parameter are

\bn{e3}
\xi^{2}_{\rm Vph} = \frac{V_{,\phi}}{V^2}V_{,\phi\phi\phi},
\ee
\bn{e4}
\sigma^{3}_{\rm Vph}=\frac{V^{2}_{,\phi}}{V^{3}}V_{,\phi\phi\phi\phi}.
\ee
Using Eqs. (\ref{e3}) and (\ref{e4}), we find $\xi^{2}_{\rm Vph}=(2/|\ga|)^{2}$ and $\sigma^{3}_{\rm Vph}=(2/|\ga|)^{3}$, respectively. For large $|\ga|$, as one could expect, the following relations hold: i-$\sigma^{3}_{\rm Vph}\ll \xi^{2}_{\rm Vph} \ll \eta_{\rm Vph}$ and  ii-$\sigma^{3}_{\rm Vph}\ll \xi^{2}_{\rm Vph} \ll \epsilon_{\rm ph}$.  In the general case, this method leads to $\lambda ^{n}=V^{n-1}_{,\phi} V^{n+1}_{,\phi}/V^{2}=(2/|\ga|)^{n}$, where the symbol $V^{n+1}_{,\phi}$ stands for the $(n+1)$-derivatives of $V$ with respect to its argument.  

We are going to discuss some heuristic issues of our model. We begin by recalling that the effective spectral index is defined as \cite{liddle2}
\bn{eindex}
n_{s}(k)-1\equiv \frac{\partial \ln {\cal P}_{s}}{\partial \ln k},
\ee
where ${\cal P}_{s}$ is the scalar power spectrum. In the case of power-law inflation  with ${\cal P}_{s} \propto k^{n_{s}-1}$,  the spectral index is a constant and is given  by $n_{s}-1=-6\epsilon_{\rm Vph}+2\eta_{\rm Vph}=2/|\ga|$. Besides, the relative error in estimating ${\cal P}_{s}$ involves the error in $|\ln k|$, typically of order $0.01$  and the error  $|n_{s}-1|<0.03$ for Planck mission \cite{planck15}. The other magnitudes in calculating $\Delta {\cal P}_{s}/ {\cal P}_{s}$ require an estimation of $n_{s}-1$ and $d n_{s}/d \ln k$ within the SC approximation. Such results can be extended from the book of Liddle and Lyth \cite{liddle2}, in fact,  neglecting higher corrections yields
\bn{eindex2}
n_{s}-1= 2/|\ga| + {\cal O}(\xi^2_{\rm Vph}),
\ee
\bn{deriva}
\frac{d n_{s}}{d \ln k}= -16\epsilon_{\rm Vph}\eta_{\rm Vph}+ 24\epsilon^2_{\rm Vph} + 2 \xi^2_{\rm V ph}= {\cal O}(\sigma^3_{\rm Vph}),
\ee
Eq. (\ref{eindex2}) tells us that to ensure some degree of accuracy of $n_{s}-1$ within the SC approach, $|\xi^2_{\rm Vph}|\ll {\max} (\epsilon_{\rm Vph},\eta_{\rm Vph})= 2/|\ga|$, whereas the slope of $n_{s}$  is zero within this approximation, so no error is introduced, as can be seen from  (\ref{deriva}).

We end this section by making some comments about the e-folding and  its connection with the amount of inflation. A useful way to address the possibility of having  super-inflation is by analyzing the number of e-folding. We need to examine if large values of $\gamma$ is consistent with  having  a sufficient  time interval to inflation takes place.  The amount of expansion during inflation is parametrized by the number of $e$-foldings $N$ which satisfies  the following relation:
\bn{ef}
dN \equiv Hdt = d \ln a.
\ee
From $(\ref{ef})$ and $\phi(t)$, we obtain $a \simeq a_{i}e^{-\lambda_{1} \epsilon_{\pm} (\phi -\phi_{0})}$. Taking $\epsilon_{\pm}=+1$ and $\phi \simeq 0$ at the end of inflation, the amount of e-folding is $\Delta N \simeq \lambda_{1}\phi_{0}$. For $\Delta N >60$ the initial field  is considerably small provided $\phi_{0} > (60/\lambda_{1}) M_{p}$, where $M_{p}$ is the Planck mass and we restored the units for sake of clarity. In another words, the end of inflation is identified with the breakdown of the slow-climb approximation. This is equivalent to $|\ga| \simeq 1$ or $\epsilon_{\rm ph} \simeq 1$ and $\delta_{\rm ph} \simeq 1 $. We can estimate the time $t_{\rm f}$ at which super-inflation ends by integrating $(\ref{ef})$ from an initial time (very close to zero) $t_{\rm i}$ to  $t_{\rm f}$. It follows that $t_{\rm f}= t_{\rm i} e^{-\Delta N/ |\gamma|}$ thanks to $H= -|\ga|t^{-1}$. Starting from a Planck era characterized by $t_{\rm i } = t_{\rm Planck} \simeq 10^{-43} {s}$, we  get the standard value  $t_{\rm f}=10^{-33}s$ by taking $|\ga| \simeq \sqrt{7}$; this is an approximate value provided  we are not considering the exact scale factor, however, the other corrections are sub-leading terms. 

\section{The phenomenology of super-inflation with phantom fields}

Here we will collect  and discuss basic mostly well-known facts about phantom cosmology, focusing  on the mechanism of super-inflation. The main idea is to evaluate in further  depth the phenomenology of the effective theory for a phantom cosmology and its implication for super-inflation. In particular, we will mention that a super-inflation model based on a phantom field  still could be justified by considering it as an effective theory within a larger scheme.    In doing so, we  will  mention  several articles which deal with the perturbations of the phantom inflation in alternative frameworks. 

The issue of phantom super-inflation was analyzed  by several authors \cite{ph1}, \cite{ph1a}-\cite{ph1e}, \cite{ph2}, \cite{ema}. To be more specific,  Piao et al. explored the issue of phantom inflation and the primordial perturbation spectrum in the case the universe exhibits  a de Sitter phase \cite{ph1}. In addition, they found  that the issue of eternal inflation could  problematic within this specific scheme and decided to include an additional normal scalar field, as is usually done in the context of the hybrid inflation model. In that paper, the authors already mentioned the possibilities of quantum instabilities when a phantom inflaton field is coupled to normal matter. Nevertheless, they considered that is much more important to focus in another aspect of the model as it may be the possibility of having a graceful exit from inflationary stage toward a radiation era. Besides, the necessary conditions to prevent  eternal inflation with phantom field were  examined by including  stochastic effects in the Langevin equation \cite{ema}, where  a white Gaussian noise is responsible for the quantum fluctuations.

At this point, it is important to recall that our model does differ from the one reported in \cite{ph1}, \cite{ph1a}, \cite{ema} provided the background is completely different in several aspects. First, the scale factor is given by a double power-law function of the cosmic time. Second, the transition from an inflationary era toward a radiation phase is achieved within the phantom model without the inclusion of further degree of freedoms.  Therefore, it is  the phantom scalar field  which decays into itself, converting the  energy stored during inflation into radiation.  Such aspects are not covered  in Refs. \cite{ph1}, \cite{ph1a}, \cite{ph2}, \cite{ema}.  Third,  flipping the sign of slow-climb parameters  associated with our model,  we recover the usual result of the spectral index for the non-phantom field, showing the consistency of our results. 

An alternative scenario to the standard inflationary model was examined by Piao et al.; they  determined the behavior of primordial perturbation spectra  with a phantom field but in the presence of expanding and contracting phase  before the bounce \cite{ph1b}.   Subsequently, Piao showed that  the phantom field  led to a very specific kind of primordial gravitational waves  which can be used for looking  new kind of physics. One of the important signature of the spin two perturbations associated with this model is the low amplitude for large scale and how it increases its magnitude  at high frequencies \cite{ph1d}. 

We mention that our work  is a clear extension of previous models based on phantom/ghost  inflation cosmology and we are considering our model as a toy one. As  was pointed out by Piao in Ref. \cite{ph1e}, the phantom scenario  can be viewed as a phenomenological framework for describing a super-accelerated early stage of the universe. For example,  phantom inflation can be accommodated  in a class of warped compactification with the brane/flux annihilation within the context of string theory.  The physical realization of this model showed the richness contained by the phantom super-inflation and why the model  is should be considered as an effective one which can be embedded  in a larger scheme, where  the phantom fields  are just one part of a fundamental  theory with extra dimensions.

The rising of phantom behavior or the existence of ghost field appears in different context of scalar-tensor theory \cite{staro} or k-essence model \cite{ke1}, \cite{kinfla}. So, our  philosophy  regarding the issue of the existence of phantom field is  described in the following. The Lagrangian associated with  the phantom field is assumed to be not an exact description of reality, but is  rather an effective  model which can be trusted at certain  energy level or energy scale. As a matter of fact,  it is widely known that the standard model of particle can only be trusted  below an energy cutoff, which is near the Tev scale.  Indeed,   Carroll, Hoffman and Trodden took the  idea of an  effective field theory as a point of departure for showing that a phantom model can be considered as a piece or a sector of a more fundamental theory \cite{carroll}.   Interestingly enough,  the aforesaid authors demonstrated that  there is  a cutoff, close to the milli-electron volt,  such that the phantom field  would be stable over the lifetime of the universe \cite{carroll}. In order to reinforce such  idea   Arkani-Hamed at al. reported  a ghost condensation within the context of an infrared modification of gravity which is  consistent and  compatible with all current experimental observations \cite{arkani}. At this point, we must stress that in our model there is a transition from an initial (inflationary) era described by a phantom field and a final era described by a quintessence-like field (see Sec. III for further details).

\section{Classical perturbations}

Observations of the CMB reveal that the universe was
nearly homogeneous at the time of decoupling. Small inhomogeneities only existed at
the $10^{−5}$ level which suggests that the conditions of the early universe may be accurately
described by means of small perturbations on top of the homogeneous background
solution. It is therefore natural to split quantities such as the metric, matter fields,  and
the stress tensor into a homogeneous background, that depends only on cosmic time,
and a small spatially dependent perturbation \cite{muka1}, \cite{muka2}.

We are interested in perturbations of the phantom field and the metric around the
homogeneous background solution which we considered in the previous sections. To be more precise, 
we are going to consider $\phi(t, x) \rightarrow \phi(t) + \delta\phi(t, x)$, $g_{\mu\nu} \rightarrow g_{\mu\nu}+\delta g_{\mu\nu}$
along with the perturbed energy-momentum $T_{\mu\nu} \rightarrow T_{\mu\nu}+\delta T_{\mu\nu}$. The dynamic of first order
scalar perturbations of the metric tensor and the energy-momentum tensor are coupled through the perturbed
Einstein equation such that a perturbation in the energy-momentum density will induce
a perturbation in the curvature of space time, $\delta G_{\mu\nu}= \delta T_{\mu\nu}$.

The metric for the background and scalar metric perturbations  can be recast as
\bn{pm}
ds^{2} = - (1 + 2\Psi) dt^2 + (1 - 2\Psi)  a^{2} (t)\delta_{ij} dx^{i} dx^{j},
\ee
where $\Psi$ is the scalar mode of the perturbed metric. In fact, the form of the perturbed metric (\ref{pm})
indicates that the anisotropic tensor vanishes. $\Psi$  is  called the curvature perturbation since 
it is the intrinsic scalar curvature of constant  time hypersurfaces, namely, $^{3}R=(4/{a^2})\nabla^{2}\Psi$.
The specific form of (\ref{pm}) tells us that we are working within the Newtonian  gauge. 
The components of perturbed energy-momentum tensor for a scalar field read \cite{muka1}, \cite{muka2}
\bn{pf1a}
\delta T^{0}_{0}=-[\Psi\dot{\phi}^2 -\dot{\phi}\dot{\delta\phi} +V_{,\phi}\delta\phi],
\ee
\bn{pf1b}
\delta T^{i}_{j}=[\Psi\dot{\phi}^2 -\dot{\phi}\dot{\delta\phi} -V_{,\phi}\delta\phi]\delta^{i}_{j},
\ee
\bn{pf2}
\delta T^{0}_{i}=\frac{1}{a}\dot{\phi}\delta\phi_{,i}~, ~~\delta T^{i}_{0}=-\frac{1}{a}\dot{\phi}\delta\phi_{,i},
\ee
where the dot stands for derivatives with respect to the cosmic time while the suffix ${, i}$ represents spatial derivatives \cite{muka2}. Taking into account (\ref{pm}), (\ref{pf1a}), (\ref{pf1b}) and (\ref{pf2}), the perturbed Einstein equations can be summarized 
\bn{peq1}
3H \dot{\Psi} + (3{H}^2 -\frac{\nabla^{2}}{a^2})\Psi= -\frac{1}{2}[\Psi\dot{\phi}^2 -\dot{\phi}\dot{\delta\phi} +V_{,\phi}\delta\phi],
\ee
\bn{peq2}
(\dot{\Psi}+  H \Psi)_{,i}=-\frac{1}{2}\dot{\phi}\delta\phi_{,i}, 
\ee
\bn{peq3}
\ddot{\Psi} + 4H\dot{\Psi} + (2\dot{H} + 3{H}^2)\Psi= \frac{1}{2}[\Psi\dot{\phi}^2 -\dot{\phi}\dot{\delta\phi} -V_{,\phi}\delta\phi].
\ee
It is important to remark that the specific form of the  perturbed Einstein equation along with the perturbed energy-momentum for the phantom field is based on the metric signature adopted, so it will  differ from the standard inflaton field equations of other  references \cite{muka1}, \cite{muka2}, \cite{cderi}.  It is useful to recall that the curvature perturbation ${\cal R}$ is a gauge invariant quantity which in the Newtonian gauge can be written as ${\cal R}= \Psi + {\cal H} v^{N}$ \cite{muka2},  $v^{N}$  being the Newtonian velocity associated with the phantom field. Using (\ref{peq2}) along with that analogy of a perfect fluid for the usual energy-momentum tensor [ i.e. $T^{0}_{~i}= -(\rho_{\phi} + P_{\phi}) v^{N}_{,i}$] one can extract $v^{N}$ and consequently  the curvature perturbation reads,
\bn{cp}
{\cal R}=\Psi + \frac{2}{3}\frac{\dot{\Psi}+  H\Psi}{ H(1+w)},
\ee
where we used that $P_{\phi}=w\rho_{\phi}$ along with the Friedmann equation ($3H^2=\rho_{\phi}$). In order to prove the constancy of this magnitude on superhorizon scale ($k\ll  aH$), we first  obtain the dynamical equation  for the gravitational potential at the  classical level. To do so,  we combine Eqs. (\ref{peq1})-(\ref{peq3})  to get
\bn{Gf1}
\ddot{\Psi} -\frac{\nabla^{2}}{a^{2}}\Psi+ \dot{\Psi} \big( H-2\frac{\ddot{\phi}}{\dot{\phi}}\big) +2\Psi \big( \dot{H}-\frac{\ddot{\phi}}{\dot{\phi}} H\big)=0. 
\ee
With the help of the  modified KG equation  ($\ddot{\phi} +3 H\dot{\phi}-V_{,\phi}=0$) along with the definition of adiabatic square speed for the phantom field  [$c^{2}_{s}=\dot{P}_{\phi}/\dot{\rho}_{\phi}$], one obtains 
\bn{speed}
c^{2}_{s}=\frac{-1}{3 H}\left(2\frac{\ddot{\phi}}{\dot{\phi}} +  3H\right).
\ee
On the other hand, the equation of state ($w=P_{\phi}/\rho_{\phi}$) satisfies the relation
\bn{eos}
3(w+1)H^2=-2\dot{H},
\ee
provided the  well-known condition (\ref{kt}) holds. Replacing (\ref{speed}) and  (\ref{eos})  into (\ref{Gf1}), the master equation for the gravitational field  yields
\bn{Gf2}
\ddot{\Psi} -\frac{\nabla^{2}}{a^{2}}\Psi+ \dot{\Psi}H \big(4+3c^{2}_{s}\big) +3H^2(c^{2}_{s}-w)\Psi=0. 
\ee
We should mention that (\ref{Gf2})   can be  used  for showing the non-evolution of the  curvature perturbation ${\cal R}$. Taking the derivative of (\ref{cp}), and employing (\ref{Gf2}), we obtain
\bn{CP2}
\dot{{\cal R}}_{k}=-\frac{k}{aH} \left[\frac{2k \Psi_{k}}{3a(1+w)} \right],
\ee
where we used a Fourier mode expansion for $\Psi=\Psi_{k}(t)e^{ikx}$ along with the relation $\dot{w}=-3H(1+w)(c^{2}_{s}-w)$. Eq. (\ref{CP2}) tells us  ${\cal R}_{k}$ remains constant on superhorizon scale for $k\ll  aH$ if the bracket remains constant or vanishes in this limit also. The aforesaid fact is a crucial point because it enables us to connect perturbations generated during inflation to perturbations after the inflation even if we do not know anything about the physical mechanisms inside the horizon during the end of inflation and reheating \cite{muka2}.  In order to achieve such a goal, we expand the perturbed quantities in Fourier modes and use the background quantities such the scale factor, the phantom field, and the Hubble parameter given in Sec. II in terms of the cosmic time.  Taking into account the previous facts Eq. (\ref{Gf1}) or Eq. (\ref{Gf2}) becomes
\bn{Gf3}
\ddot{\Psi}_{k} +\frac{(2-|\ga|)}{t}\dot{\Psi}_{k}+ \frac{\bar{k}^2}{t^{-2|\ga|}} \Psi_{k}=0.
\ee
Here $\bar{k}=k/a_{i}$. Eq. (\ref{Gf3}) admits a closed solution in terms of a linear combination of the base  ${t}^{q} \{J_{p}(nx),  J_{-p}(nx)\}$ where $x=(\bar{k}^{1/(1+|\ga|)}t)^{(|\ga|-1)/2}$,  $q=(|\ga|-1)/2<0$, $p=(|\ga|-1)/2(|\ga|+1)>0$ and $n$ some constant which can be cast in terms of $|\ga|$. We are not going to list that result here, however,  there is a way to reduce (\ref{Gf3}) into the standard Bessel equation form by applying just a simple parametrization of the solution as $\Psi=t^{m}P(t)$ and choosing $m$ properly  (see \cite{me} for further details).  We want to determine  the form of the gravitational field when the cosmic time is  very close to zero, thus we only need to capture the leading behavior of  $\Psi_{k}$. Implementing the previous approach into Eq. (\ref{Gf3}), the gravitational field reads
 
\bn{GfS}
{\Psi}_{k}(t)=C_{1k}t^{\frac{(1-|\ga|)}{2}}+ C_{2k} t^{\frac{|\ga|-1}{2}}
\ee
where $C_{1k}$ and $C_{2k}$ are integration constants. It is clear from (\ref{GfS}) that the gravitational field has two kinds of modes, a growing one and a decaying mode. Let us  explore  the physical consequences of coming from the leading term  in Eq. (\ref{GfS}). In the case of  small $k/(aH) \propto (k/a_{i}) t^{(1+|\ga|)}$, the derivative of the curvature perturbation (\ref{CP2}) tells us that
\bn{RCur}
\dot{{\cal R}}_{k}\simeq  -\frac{2k^{2}}{3|\ga|(1+w) a^{2}_{i}}t^{3\frac{(1+|\ga|)}{2}} \rightarrow 0
\ee
Bearing in mind all this information we need to find the behavior of the perturbed phantom field during inflation. One way to obtain  $\delta \phi_{k}$ is by looking more closely to the perturbed Einstein constraint (\ref{peq1}). It is important to note that the evolution of background quantities (namely, $\phi$, $V_{,\phi}$ and $H$) and the gravitational field $\Psi_{k}$ are known at this stage. Therefore, the perturbed Einstein constraint can be seen as master equation for $\delta \phi_{k}$. Of course, all the physical results that we can extract from this equation are equivalent to those  obtained from (\ref{peq2}) or (\ref{peq3}). We  present (\ref{peq1}) as  a differential equation for $\delta\phi_{k}$
\bn{peq1B}
\dot{\delta\phi}_{k} -\frac{\dot{V}}{\dot{\phi}^2}{\delta\phi}_{k}= S_{k}(t),
\ee
and the source term is given by  
\bn{ST}
S_{k}(t)=\frac{{\Psi}_{k}}{\dot{\phi}}\left(\frac{2k^2}{a^{2}}+\dot{\phi}^2\right)+ \frac{6H^2}{\dot{\phi}}\Psi_{k}+ \frac{6H}{\dot{\phi}}\dot{\Psi}_{k}.
\ee
Before solving (\ref{peq1B}),  it is illustrative  to  determine the leading contribution of each term in (\ref{ST}). To be more precise, we obtain  
${\Psi}_{k}/a^{2}\dot{\phi} \propto t^{3(1+|\ga|)/2}$,  $ {\Psi}_{k}\dot{\phi} \propto t^{-(1+|\ga|)/2}$, $ \dot{{\Psi}}_{k}H/\dot{\phi} \propto t^{-(1+|\ga|)/2}$, and $ {\Psi}_{k}H^2/\dot{\phi}  \propto t^{-(1+|\ga|)/2}$. In short, the leading contribution in  (\ref{ST}) can be written as $S_{k}(t)=S_{0} t^{-(1+|\ga|)/2}$ with $S_{0}$ a constant. Replacing the latter result along with  $\dot{V}/\dot{\phi}^2=-(3|\ga|+1)t^{-1}$  in (\ref{peq1B}), we see that the perturbed field is 
\bn{perf}
{\delta\phi}_{k}(t)=D_{1k}t^{-(3|\ga|+1)}- \frac{2S_{0}}{(7+5|\ga|)} t^{(1-|\ga|)/2}.
\ee
Both terms in Eq. (\ref{perf}) are growing modes in the limit of vanishing cosmic time when phantom inflation happens. However, the leading contribution is due  to the homogeneous solution of  (\ref{peq1B}). Eq. (\ref{perf}) shows us that the perturbed (phantom) field grows without limit near inflation but it then decays at  late times. Another physical quantity that we must calculate is the contrast density, at least its leading contribution. The evolution of $\delta \rho_{\phi}$ can be obtained from (\ref{pf1a}) using  $\delta T^{0}_{0}=-\delta \rho_{\phi}$ along with (\ref{GfS}) plus (\ref{perf}).
It turns out that   $\delta \rho_{\phi}$  has three contributions, namely, $\Psi\dot{\phi}^2 \propto t^{-(3+|\ga|)/2}$, $\dot{\phi}\dot{\delta\phi}\propto t^{-(3|\ga|+3)}$, and  $V_{,\phi}\delta\phi \propto   t^{-3(1+|\ga|)}$.  From the latter fact and the background energy density (\ref{dt}), we find that the density contrast  behaves as
\bn{cod}
\frac{\delta \rho_{\phi}}{\rho_{\phi}} \simeq \frac{\delta\rho_{i}}{3\ga^{2}\rho_{i}} t^{-3(|\ga|+1)},
\ee
where $\delta\rho_{i}$ is an initial density contrast. As one might expect, Eq. (\ref{cod}) tells us that the density contrast grows like  the perturbed  phantom field, thus ${\delta\phi}_{k}(t) \propto t^{-3(|\ga|+1)}$.   

Of course, the analysis  is not complete and it is important to establish what happens in the opposite limit, during a matter-like era. Following the same approach for examining the classical perturbation during a matter(radiation)-like era (the precise kind of era is controlling by the value of $\beta$), we obtain similar equations for the gravitational potential and therefore we are not going to repeat the procedure here.   We find that the growing mode of gravitational field is ${\Psi}_{k}\propto t^{(\beta+1)/2}$.  We have two different types of behaviors for the perturbed field depending on the values which are  taken by $\beta$. For $\beta \in (0, 7/9)$, we find that the leading term in the perturbed field is  $\propto  t^{(-3\beta +5)/2}$ and  density contrast  is given by  $ (\delta \rho_{\phi}/\rho_{\phi}) \propto t^{(-3\beta+5)/2} $ also;  the growing behavior of the density contrast at late time ensures the formation of structures. For  $7/9<\beta<1$,  the matter  grows  at a different rate, namely $(\delta\rho_{\phi}/\rho_{\phi}) \propto {\delta\phi}_{k}(t) \propto  t^{(3\beta -1)}$. 

\section{Conclusions}
In this paper we have analyzed  one transient model within the context of ``phantom'' inflation. We presented an exact solution where the scale factor interpolates between a pole-like inflation era and  a radiation era at late times. Given a scale factor,  we have reconstructed the potential energy and kinetic in terms of the cosmic time, further we found that the potential corresponds to an exponential one. An interesting fact of this model is that the phantom inflaton decays into a quintessence field converting the stored energy during inflation into coherent radiation without the need of an extra field to produce such mechanics. Besides, we showed that the phantom model is free of the eternal inflation issue.  We examined the issue of a phantom inflationary model that crosses the divide line before the universe enters the non-accelerated second stage. We also showed that the original phantom  Lagrangian exhibited a global symmetry which allowed us to explained the reason why the model could describe the radiation (or matter-like) final era.

We have examined  the  slow-climb approximation  and determined the conditions under which the model can be considered as a viable inflationary  scenario. The analysis was carried out  with the help of the (kinematic) slow-climb parameters, showing that the parameter of the model must be considerable.   We also  went in the analysis beyond the slow-climb approximation by including the slow-climb parameters in terms of the potential. In that case, we calculated the  effective spectral index and the running of it. Later, we analyzed the necessary conditions to  have a suffcient amount of  inflation. 

We pointed out the difference among our model and other interesting proposals.  The background dynamic is characterized by a scale factor  which has a double power-law function of the cosmic time. The aforesaid trait allows the transition from an inflationary era toward a radiation phase within the phantom model without the inclusion of further degree of freedoms; such aspects are not covered  in Refs. \cite{ph1}, \cite{ph1a}. 

At the level of classical perturbations we have obtained the growing and decaying mode of the gravitational field within the Newtonian gauge. We also have verified  that the derivative of the curvature perturbation goes to zero for small $k/aH$. We  have calculated the leading behavior of perturbed field during inflation; it turns that the perturbed field grows like the contrast density and both behave as a power law (pole)  with the cosmic time  during inflation and then decays to zero at late times. In addition, the perturbed field and the contrast density both grow at late time. 

Summarizing, unlike  some previous work on phantom inflation (cf. \cite{ph1})  where an extra scalar field is introduced to force the phantom field to decay,
our model  only has one phantom inflaton  that decays into another quintessence field, such a field being  responsible for both the inflationary epoch and the radiation era. 

\acknowledgments
We are grateful to Prof. Dr.  A. Vikman for many useful comments on the article and the reviewer for some important suggestions.
M. G. R. and G. M. K. are supported by Conselho Nacional de Desenvolvimento Cient\'ifico e Tecnol\'ogico (CNPq)- Brazil.


\begin{thebibliography}{99}
\bibitem{tech}
An issue that could be problematic is the quantum field theory associated with a phantom field.
For instance, the model could lead to an instability of a vacuum state, however, one can always
consider this field as an effective  realization of a fundamental non-phantom theory well formulated
in terms of the standard scheme of QFT. For a useful discussion of this point, the reader may see the references
mentioned in \cite{ema}.
\bibitem{f}
We have dropped the $a_{0}$ factor for sake of simplicity. However, they can be easily restored by re-scaling $a$ by $\frac{a}{a_{i}}$. The same kind of procedure
can be done for the energy density, pressure, kinetic energy and potential energy; for instance $p \rightarrow p/p_{i}$,  $V \rightarrow V/V_{i}$, etc. We will consider the physical case in which $a_{i}>0$, $p_{i}>0$, $V_{i}>0$, and $\rho_{i}>0$.
\bibitem{f2}
The second scenario is characterized by $ H<0$, $\dot{H}>0$, and $\frac{\ddot{a}}{a}>0$. Therefore,  the Ricci curvature is positive defined
provided $R= 6[\dot{H} + 2{ H}^2]$.
\bibitem{cderi}
Nevertheless, our derivations are consistent with our previous choices made in the variational principle for obtaining the Einstein's field equations.  From  our action (\ref{a}) and applying the variational principle, we arrive at $G_{\mu\nu}=+T_{\mu\nu}$. Also,  the specific form of the perturbed energy-momentum is consistent with the adopted Lagrangian for the phantom field, as one can easily check at the background level by using  $L_{\phi}=X-V$ and the definition of $T^{\phi}_{\mu\nu}=-\frac{2}{\sqrt{-g}}\frac{\delta {\cal L}^{\phi}}{\delta g^{\mu\nu}}=-{\cal L}^{\phi}_{,X}\partial_{\mu}\phi\partial_{\nu}\phi + g_{\mu\nu} {\cal L}^{\phi}.$ In order to compare our results with previous works \cite{muka2}, some cautions must be taken in converting them and adapted to other frames. For instance, the energy-momentum of the perfect fluid is $T^{\mu}_{\nu}=(\rho+P)u^{\mu}u_{\nu}+P\delta^{\mu}_{\nu}$ while the constraint for four-vector is given by $u^{\mu}u_{\mu}=-1$.
\end{thebibliography}
\end{document}